# Spike train propagation in Hybrid Artificial Neural Network (HANN).

## Contoyiannis. F. Yiannis[1].

1. Department of Electrical and Electronics Engineering, University of West Attica, Ancient Olive Grove Campus, 250 Thivon and P. Ralli, Athens GR12244, Greece.

   *Correspondence: yiaconto@uniwa.gr

**Abstract**: The spikes train is an important step in order to the artificial neural network (ANN) give us simulations more close to the reality i.e the operation of the biological neural network. Based on in previous our work that the HANN can to produce critical and tricritical intermittencies we investigate in present work the possibility of the Spike train production from the HANN. So the operation of ANN does not would based in mathematical algorithm of machine learning but the operation of a ANN could be based in physical notions as the phenomenon of intermittency. As we have shown the real biological neurons is a Dynamical system which present the intermittent dynamic type I.

# Section1 : Introduction

Today's ANNs have an important application in the production of artificial intelligence (AI). The concepts borrowed from the natural intelligence of living beings, namely the functioning of the brain, are basically two. The concept of the neural network and the concept of the synapses where the connection of neurons is made. Their design is based on mathematical algorithms that utilize fast computers and huge databases. However, the Physical-Biological reality is very far away. The most basic of all is missing, which is the dissemination of information on the network. In biological neurons, this is done through the axion potential, i.e. the production of spike train (ST), which is the temporal evolution of spike structures that propagate through the neuron axons and are transmitted to the network through the synapses. The most important attempt to simulate the biological ST has been made with the pioneering work of *Hodgkin AL, Huxley AF* [1]. There, a system of differential equations generated by an electronic circuit that simulates the production of the membrane potential reproduces the shape of the spike of the biological neuron, with the main characteristics being the threshold of the potential and the phenomenon of hyperpolarization. In our works [2,3] on biological ST, we have found that in addition to the two aforementioned characteristics of biological spikes, there are and other significant characteristics, which are the critical dynamics to which the ST time series obeys during the intervals of the relaxation phase (ST grass) as well as the fractal structure in the inter-spike intervals [3].

The present work is an application of the recently introduced mechanism where we produced the main characteristics of the biological membrane potential i.e the form of spikes ( threshold and hyperpolarization) , the fractal structure of inter-spikes intervals as well as the critical dynamics of the fluctuations in the grass region where the relaxation temporal interval between the spikes appears. This mechanism is based on a coupling procedure between of two types intermittent maps . These maps express mathematicality the dynamics of the order parameter fluctuations in critical point of a second order phase transition and the dynamics of the order parameter fluctuations in the begging of a crossover procedure around the Griffiths tricritical point which is the connection point between the second and first order phase transitions in parametric space. Due to the universal character of the intemittency phenomenon this mechanism can to applied in all the dynamic systems real or numerical . A necessary prerequisite is the time series that express the temporal evolution of the order parameter ( time series) to obey both dynamics namely the critical and tricritical intermittencies, so that their coupling is possible. Such a system is the HANN for which the application of spike production like the biological spike train based on this coupling is the object of this work. As we said before the upgrade of an HANN towards the biological nature should also use moreover and other biological concepts such as the action potential, i.e. the production of spikes, and more generally the membrane potential, which also includes the fluctuations of the potential in the relaxation time intervals (grass). In recent years, such an effort has begun. A ANN that fires at the moment of threshold crossing is called spiking neural network (SNN) [4-9] The most prominent SSN are the -fire models [10,11] In these models, the activation level is considered to be the neuron's state where, if the firing threshold is reached the neuron fires. After firing, the state variable is reset to a lower value ( like hyperpolarization phenomenon). Spike-based activation of SNNs is not differentiable thus making it hard to develop gradient descent based training methods to perform error backpropagation [12,13]. One way to overcome this problem is in case where the sigmoid function could approximates the Heavyside step function which is differentiable [3 ] . Various techniques have been developed in order to equips SNNs with improved learning capabilities, elevates computational efficiency[14-17] . However, there is a lack of learning mechanisms for SNNs, which can be inhibitory for some applications. In a recent work [18], we overcame this problem by producing STs with a completely different mechanism that is based on the coupling of two types of intermittency. The result was to approach the biological neuron both qualitatively and quantitatively very closely. In the present work, we apply this mechanism to time series produced by HANN that obey the two forms of intermittency. This will open new avenues for ANNs used in AI by enriching it with basic principles of Physics and Biology, thus bringing it closer to Nature.

## Section 2:  The phenomenon of intermittency

The phenomenon of intermittency is the temporal alternation of regions of low fluctuations called laminar regions  with  regions  which demonstrate  high fluctuations  ( bursts) .

 In figure  1   such  a timeseries  which present this alternation  is shown. The way  to construct a numerical simulation through the intermittent map ( eq.1)  to produce  an intermittency time series as in figure 1 is to find its successive values through the following way : That is, as soon as an burst ends, the trajectory  returning randomly to the  values of the laminar region, e.g. [0,0.2] from fig.1   and so on.  Considering the laminar region as the input values then the bursts    are the output values .  So, the return from output   back  to  input  of the laminar regions values i.e  a feedback character  in this procedure exist  . Therefore, the backpropagation which is a feedback mechanism used to optimize the error is accomplished  through the feedback mechanism of intermittency that Nature uses in many phenomena. Moreover the universality of intermittency is a prerequisite for the emergence of an universal  mechanism we are looking for  the production of the spikes.

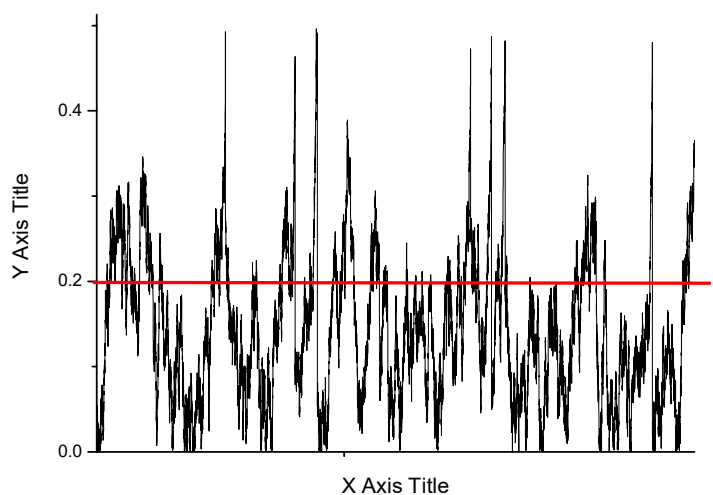

Fig1. *An intermittent timeseries where we have suppose that the zone from 0 up to 0.2 ( red line ) is the laminar region and every value which is upper from this zone belong in bursts*

The laminar lengths L are the waiting times inside the laminar region zone of fig1 . Thus the laminar lengths determined through the relation

$$\Phi_{red} < \Phi < \Phi_{blue}$$

The level $\Phi_{red}$ has a constant value which is the lower value of grass. The $\Phi_{blue}$ value is a free parameter. So, a sweep of grass values inside this zone is accomplished. The sweeps are finished when we find the best power-law distribution of laminar lengths It is known [19] that the critical state in the second order phase transition phenomenon obey to the critical intermittency expressed mathematically with the intermittent map type I.

$$\phi_{n+1} = \phi_n + u_1\phi_n^z + \varepsilon_n \quad (1)$$

$u_1 > 0, \ z > 0$ , $\varepsilon_n$ uniform noise between $[-\varepsilon_1, +\varepsilon_1]$.

As we have shown in our work [19] the quantities that appear in equation 1 are critical quantities. Thus as $\phi_n$ we denote the fluctuations of the order parameter and the exponent z is connected to the isotherm critical exponent δ with the relation

$$z = \delta + 1. \quad (2)$$

The waveform of a spike includes an rise and a fall. Map of eq.1 assures us of the rise, but what about the fall? It is known from the theory of critical phenomena [20] that there is a region where a crossover from the second-order phase transition to the first-order phase transition takes place. . This crossover is accomplished around the so-called Griffiths tricritical point. As we have shown in our work [21] , the dynamics in the begging of this crossover is described by a mathematical intermittent map given by relation 3.

$$\phi_{n+1} = \phi_n - u_2\phi_n^{-z} + \varepsilon_n \quad (3)$$

$u_2 > 0$, , $\varepsilon_n$ uniform noise between $[-\varepsilon_2, +\varepsilon_2]$. The negative sign of the non-linear term and the negative exponent ensures that the values fall. The two maps are repellors where the fixed point in map 1 is the zero and lead the trajectory at higher values ( rise) while the fixed point in map (eq3) is high value (theoretical the infinity ) and lead the trajectory at smaller values ( fall). The dynamics is determined by the distribution of the laminar lengths L of the intermittency. According to reference [22] for the map1 this distribution is given by the power-law :

$$P(L) \sim L^{-p} \quad (4)$$

Where $p = \frac{z}{z-1}$. Due to eq . 2 we have that :

$$p = 1 + 1/\delta \quad (5)$$

Given the fact that $1 < \delta < \infty$ from equation 5 results that the critical intermittency exist for exponent $p\epsilon[1,2]$. According to reference [21] for the map2 this distribution is given by the eq. 4 where now $p = \frac{z}{z+1}$.. . Due to eq . 2 we have that in tricritical dynamics

$$p = \frac{\delta+1}{\delta+2} \quad (6).$$

Given the fact that $1 < \delta < \infty$ from equation 6 results that the tricritical intermittency exist for exponent $p\epsilon[0.66,1]$.

$$f(L) \sim L^{-p_2} e^{-p_3 L} \quad (7)$$

The above is the Method of critical fluctuations (MCF )[23] with which we reveal how close the distribution of laminar lengths (generally waiting times) is to a power law. If we find an exponent $p \epsilon [1,2]$ then the system which produce the time series is at critical state too. ,The fitting function which we use in order to calculate the exponent of power-law is :

As we see we have introduce in eq.7 an exponential corrective term where the exponent $p_3$ appears. When this exponent is close to zero the distribution reach to the scaling form of eq.4. Thus the exponent p3 is a :measure" how close we are in criticality if simultaneously the $p2 \in [1,2]$.

## Section 3.   A brief introduction in HANN

A first step to this direction we have make in our work [3]  where we produce  a timeseries   from a HANN  model  using the notions from biological neurons networks.   In reality this model produced a  Kink-antikink  solitons  times  series  which  is  very  close  to   a Heaviside   step  functions  timeseries.  It  is  known   that    the derivative  of  the  Heaviside  step  function  is   the  Dirac  delta function  which has a form similar to a spike . This way bypasses the problem of non-derivative which is the main problem of SNN algorithms  where  learning  process  cannot  be  done  by backpropagation  methods . This methodology can produce a few spikes but most of the time series remains step functions. So our next step was to apply the coupling of the two maps  to a HANN which,  as  we  will  see  in   the  next  sections ,  also  displays  the dynamics of these maps..

In the following, we present the characteristics and the properties of the  HANN model [24] .

- We have considered N neurons that are in the states {-1,+1} without a particular topology (for example lattices or layers) . Considering the neurons as a field of neurons with values from {-1,+1} we calculate the mean value of the neurons  field .

- We follow the principle of least energy (Physical principle which for equal time intervals is equivalent  with the principle of least action) to produce successive configurations  of the neurons  field. Computationally, we do this with the Metropolis  algorithm [25], as in the Isings models.

- In the Metropolis algorithm executed flips between the states {-1,+1} . For the selection of the flip result  classically the Boltzmann statistic is used while in our hybrid model we use the Fermi statistic  to emphasize the fermionic character of the {-1,+1} states.

- For the quantity of "Energy" we use the exact definition of Energy in Physics and not the propagation function which is commonly used in ANN.
- We use biological rules to select statistical weights.

The quantities used in the model are:

- The energy function representing the state of the HANN at time $t$ is given by [24].

$$E(t) = -\sum_{i,j=1}^{n} T_{ij}\,\xi_i(t)\xi_j(t) \quad (8)$$

Where the connection weights $T_{ij}$ may take either positive or negative values, reflecting synaptic properties in the connection between two biological neurons. For network of $n$ neurons, whose output states are random variables $\xi_i, i=1,\ldots n$ that can take the values +1 or -1.

- For the elements of matrix $T_{ij}$ i, j = 1, ... n we assign the following properties:

$T_{ij} = 0$ when i=j (no self-interaction)

$T_{ij} \neq T_{ji}$ because this is closer to biological reality

Neurophysiological data suggest that excitation and inhibition are in balance in cortical networks [26]. Although on average, about 80% of synapses are excitatory, inhibition is more effective in shunting a neuron than excitation in making a neuron fire.

- The probability for the output of neuron $i$ at time $t+1$ to make a transition between the states $\pm 1 \rightarrow \mp 1$ is calculated by [24]:

$$\Pr\left(\xi_i\left(t+1\right)=+1\right)=\frac{1}{1+e^{-\beta E(t)}}$$

(9a)

$$\Pr\left(\xi_i\left(t+1\right)=-1\right)=1-\Pr\left(\xi_i\left(t+1\right)=+1\right)=\frac{1}{1+e^{\beta E(t)}}$$

(9b)

In the current model, the weight function in the Metropolis algorithm has the form of the Fermi-Dirac distribution for fermionic degrees of freedom:

$$f = \frac{1}{1 + e^{-\beta \varepsilon}}$$

(10)

which describes the probability of quasi-particle excitations in a Fermi liquid [27].

- In Eq. (10) $\beta$ is the inverse of temperature [11] and $\varepsilon$, which is usually the energy of a single fermion, expresses the energy of the entire system. A local field of this transition could be $m_i(t)$, which under the consideration $\beta_i = \beta$ takes the same value for all neurons [24]:

$$m_i(t) = \Pr\left(\xi_i(t+1) = +1\right) - \Pr\left(\xi_i(t+1) = -1\right) = \frac{1}{1 + e^{-\beta E(t)}} - \frac{1}{1 + e^{\beta E(t)}} = \tanh\left(\frac{\beta}{2} E(t)\right)$$

(11)

- We estimate the *mean field* of all neurons as:

$$Field(t) = \frac{\sum_{i=1}^{n} m_i(t)}{n}$$

(12)

Note that under the consideration that for all neurons $\beta_i = \beta$, the mean field is essentially equal to the local field. However, for purposes of generality of the algorithm we will keep the abovementioned definition of the mean field.

- We estimate the quantities $FS^{(+)}$ and $FS^{(-)}$ after a number of iterations $N_{iter}$ as $FS^{(+)} = \sum_{t=1}^{N_{iter}} Field^{(+)}(t)$, $FS^{(-)} = \sum_{t=1}^{N_{iter}} Field^{(-)}(t)$, where $Field^{(+)}(t)$ and $Field^{(-)}(t)$ are the positive and negative values of mean field time series, respectively.

- We define the *effective order parameter* as:

$$M = \frac{FS^{(+)} - \left| FS^{(-)} \right|}{FS^{(+)} + \left| FS^{(-)} \right|} \qquad (13)$$

A success of this model was that it managed to numerically calculate critical exponents of the 2D-Ising model, which had only been done analytically until then. For more details of the HNN model, please refer to our work.

In our work [3] we produce a timeseries from a HANN model using the notions from biological neurons networks. In reality this model produced a Kink-antikink solitons times series which is very close to a Heaviside step functions timeseries. It is known that the derivative of the Heaviside step function is the Dirac delta function which has a form similar to a spike . This way bypasses the problem of non-derivative which is the main problem of SNN algorithms where learning process cannot be done by backpropagation methods . This methodology can produce a few spikes but most of the time series remains step functions. So our next step was to apply the coupling of the two maps to a HNN which, as we will see in the next sections , also displays the dynamics of these maps..

## Section 4.  The intermittency phenomenon in HANN

Following the flow of equations 8, 11,12,13 we calculate the values of the time-series Field(t).  Every time we change the initial conditions determined by the initial values of the neurons we will find a different value for the effective order parameter M ( eq.13).  If we select as temperature T=2.3 (β=0, 4348 ) which is very close to 2.266 which is the critical temperature in 2D-ising model [28], then the number of the M values which are close to zero grows up.  As the value of the order parameter M is closer to zero, we are closer to criticality.

In HANN we had found an important property. This finding is that with a change of the initial conditions for the critical temperature T=2.3 and where the effective order parameter M takes small values close to zero, both types of dynamics described by the critical map (eq.1) and the tricritical map (eq.3) appear. Indeed, as very detailed in our work [24] we have shown that in HANN time series the distribution of laminar lengths is a power law with exponent p given by the eq.5 where δ=15 which is the critical isotherm exponent for the universality of the 2D-Ising model [20]. On the other hand, for other initial conditions but in the same critical temperature , a HNN time series is obtained, where the distribution of laminar lengths is a power law with exponent p given by the eq. 6 where δ=15 . Namely , now the HANN gives an result for the laminar length distribution that obey to the tricritical intermittency . In following we produce from HANN two time series that obey the dynamics of intermittencies.

In a HANN there is symmetry of the values with respect to the axis that passes through zero [24]. Therefore, we can choose positive or negative values. Thus, after cutting off the negative values, we create a new time series that appear in figure 2 (see below) for the temperature T=2.3. The time series have a length of 120000 points and have resulted from two different initial conditions which are for the fig. 6α  M=3.7 $10^{-2}$ and for the fig. 6c  M= 1.4 $10^{-2}$. We apply the MCF ( section 2) for the calculation of the laminar lengths distribution. In figs 2b. 2d these distributions are shown. With red lines in these plots are the fitting function which is given by the eq. (7). The values of the exponents for the distribution 2b which corresponds to time series  2a are $p_2$ = 1.08 ∈ [1,2) (critical state) and $p_3$=0.007 (very close to zero). These values results for laminar region [$\Phi_{blue}$ =0.7, $\Phi_{red}$ = 1] ( the zone between the red and blue lines in Fig2a). With similar way we find that the values of the exponents for the distribution 2d which corresponds to time series  2c are $p_2$ = 0.96 ∈ [0.66,1) ( tricritical state).   and $p_3$=0.02 close to zero. These values results for laminar region [$\Phi_{blue}$ =0.6, $\Phi_{red}$ = 1]  ( the zone between the red and blue lines in Fig2c).

Due to the small values of  $p_3$  we have that both distributions are very close to power-laws.  If in the equation (5) we put δ=15  for the universality class of  a D-Ising model we find the theoretical value  $p_{(2}$

Theor)=1.066 very close to 1.08. Therefore the HANN timeseries of fig.2a obey to critical intermittency (i.e map1). . Now if in the equation (6) we put δ=15 for the universality class of 2 D-Ising model we find the theoretical value p(2 Theor)=0.94 very close to 0.96. Therefore the HANN timeseries of fig.2c obey to tri critical intermittency (i.e map2). . Thus we produce from HNN two time series that obey the dynamics of the two intermittencies.

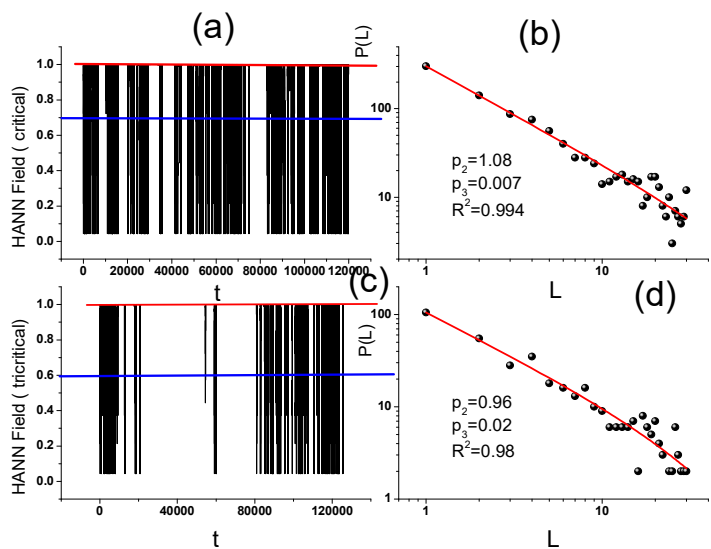

**Fig2 . (a) Critical HANN with laminar region [0.7,1]. (b) The laminar lengths distribution of Fig.2a. The exponents are: p₂=1.08 ∈ [1,2) p₃ =0.007. (c) tricritical HANN with laminar region [0.6,1]. (d) The laminar lengths distribution of Fig.2c. The exponents are: p₂=0,96 ∈ [0.66,1) p₃=0.02.**

## Section 5. The coupling between critical –tricritical HANN

In our work [18], the method of coupling of the two mathematical maps (CM) is presented analytically through the code that we presented in CM [18] . With the appropriate combinations of the parameter values, we managed to produce the biological  ST.  However, how will this coupling between the time series of  HANN (figs 2a, 2c ) be carried out?  The distribution of the laminar  lengths as we saw before ,  follows the dynamics of the maps  eq1, eq3 ,  but now here in the numerical as well as in real signals  i.e timeseries does not appears  mathematical equations  like the eq1,  eq3  with a direct way.  In real world or in numerical systems  it is possible that the dynamics ,expressed through the distribution of laminar lengths,  are the same as the dynamics of the mathematical maps , but here  cannot calculate the parameters of the equations (1,3) due to the random  processes that exist. Thus  the real world  cannot be expressed with mathematical  maps  such as equations (1,3).  Thefore,  we must modified  the  algorithm for the coupling  which applied in  the previous work [18]. However, the goal  is the same i.e one time series to create the rise  and the other to create the fall . In the maps case , the fall  is created due to the negative sign of the non-linear term of equation 3.. Given that both time series in figure 2  have values  below of 1 , we will change the sign  in a time series . Thus  the coupling will be done  as ( laminar time series1↔-laminar time series2  ).  In the case of HNNs as well as in real time series, the code parameters are the 4 values of the boundaries  Φ_blue  , Φ_red  of the laminar regions as they were calculated using the MCF method in Figure 2

**The   code for the coupling**

c      ****Production Spikes  from the HANN models****

   dimension x(500000)

   dimension y(500000)

   dimension tms(500000)

```fortran
      dimension coupltms(500000)
c     *************************************
      open(9,file="coupltms.dat")
      open(13,file="segment.dat")

c        ********Introduction data  of  HANN critical***************
         open(12,file="C:\contoyiannis\Spikes HANN\HANN tms1.for")
            do i=1,120000
                     read(12,*) x(i)
                    write(10,20)x(i)
          enddo
           close(12)
c        ********Introduction data  of  HANN tricritical*********
      open(14,file="C:\contoyiannis\Spikes HANN\ HANNtms2.for")
      do j=1,120000
            read(14,*) y(j)
                  write(11,20)y(j)
       enddo
       close(14)
c     *********Laminar region of  critical tms*************
         do i=1,120000
                     if(x(i).le.1.and.x(i).ge.0.7)then
                         tms(i)=x(i)
                          endif
         enddo
 c     *********Laminar region of  - tricritical tms*************8
         do  j=1,120000
```

```fortran
            if(y(j).le.1.and. y(j).ge.0.6)then
                    k=1
                      tms(j+k)=-y(j)
               endif
        enddo
c        *****Addition in the  grass  noise**************
        a=-0.035
         b=0.035
        do l=1,120000
                call random(r)
                    e=((b-a)*r+a)
                          tms(l)=tms(l)+e
                coupltms(l)=tms(l)
c        ********Final result****************
            write(9,40)l,coupltms(l)

        enddo
c      ************segment*********
        do i=105965,106100

           write(13,40)i,coupltms(i)
          enddo
  10  format(i8)
  20  format(f14.8)
  30   format(f14.8,5x,f14.8)
  40   format(i8,5x,f14.8)
     close(9)
```

```
close(13)

stop

end
```

The coupling is performed by mixing the two laminar regions through the line of code :

$$k=1$$

$$tms(j+k)=-y(j)$$

where  k =1,2,3,4……

## Section 6.   The   Results

We apply  the  above code  for  intermittencies  time series  from 120000 points.  The laminar region  of critical intermittency is  [$\Phi_{blue}$ =0.7, $\Phi_{red} = 1$]  and the corresponding laminar region of tricritical intermittency is  [$\Phi_{blue}$ =0.6, $\Phi_{red} = 1$]  .  In the grass of spikes train a uniform noise  has been added.  Unlike the  our work  where  the noise already exist in the equations 1, 3   here  we add  the  noise with width [-0.035, 0.035]  which is the same with noise of HANN in our work   [24]. As we have said before, the appropriate noise helps to appear  more clear the critical fluctuations if these  really exist in the signal.  For the coupling of two laminar regions  we  put as k=1.  In fig 3 we present the results from a   segment of length 11000 points  from the 120000  points for clarity reasons  of figure.

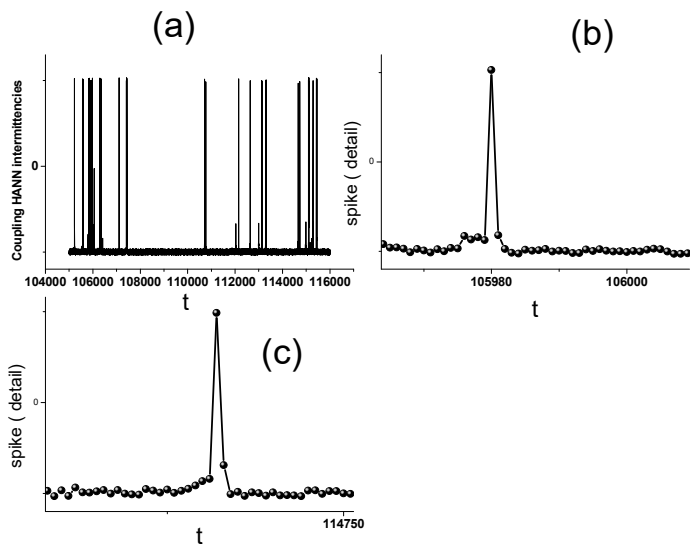

**FIG.3(a)  A  segment of length 11000 points  produced from the coupling  between the laminar region  of  critical  intermittency ( fig2a)  and the laminar region  of tricritical  intermittency  ( fig 2c). (b,c)  Details of the two characteristic  spikes from the  spike train of fig.3(a).  The characteristics (   hyperpolarization phenomenon   and  threshold ) similar to   biological spikes  are  obvious.**

In next  figures  4,5  we present, for comparison reasons ,   Spikes Train from   : (a)  biological membrane potential (b)   Coupling of maps (CM) and (c)  from  HANN model.

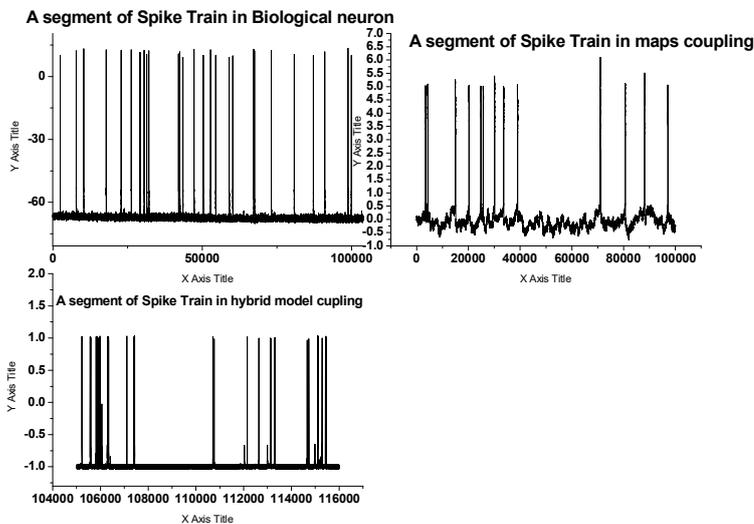

Fig4. (a) *)* *A segment of 103000 point from biological membrane potential from in vitro intracellular recordings of CA1 pyramidal neurons in Wistar male rats.* (b) *A segment of 100000 point from coupling timeseries in Intermittencies coupling mechanism (CM) (c) A segment of 11000 point from coupling time series in Hybrid model.*

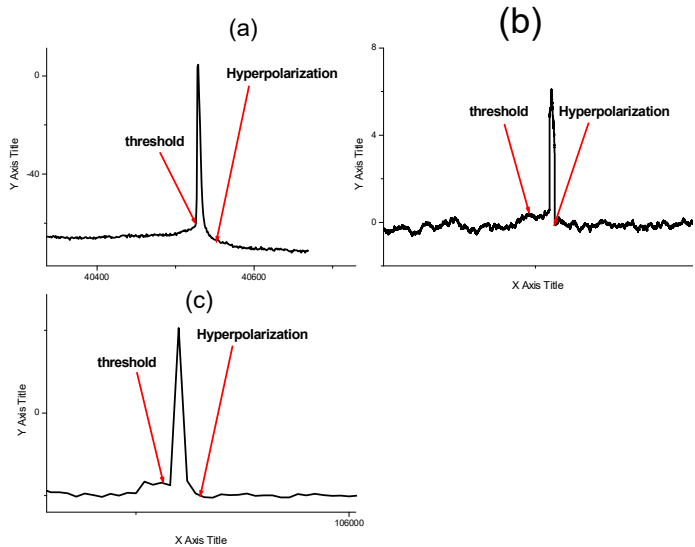

Fig .5 (a) *A spike from biological membrane potential* (b) *A spike from CM model (c) A spike from HANN model.*

The next step is to examine whether the grass obey to critical dynamics as we have already seen in the biological neurons [2] and in the CM model [18] . For this purpose, we will apply the MCF method to calculate the distribution of laminar lengths in the laminar region of the grass. If this distribution is a power law with exponent p [1,2] then we can really talk about the production through the coupling mechanism of a biological type spike signal. We apply the MCF in the whole length of timeseries i. .e 120000 points because we want the best Statistics. However, for reasons of better appearance of the grass area in figure 6 we show the grass of the segment of fig.4c. Looking for the laminar

**18**

region that gives the closest result to a power law we found that the laminar region is the zone [-1.035, 0.948] as seen in figure 6a.

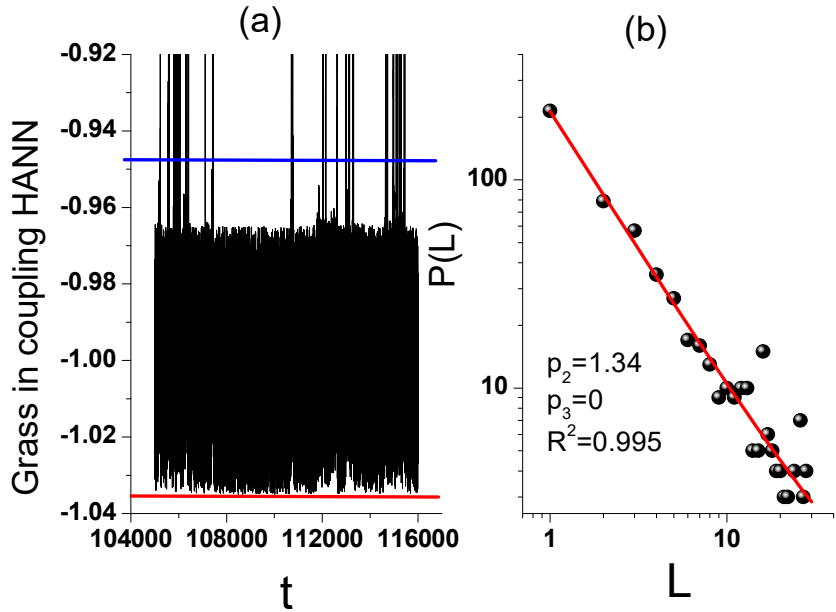

*Fig.6 (a) The laminar region in the grass in the effective intermittency which result from the coupling of the critical and tricritical intermittencies (fig 2a, fig2c ). This grass laminar region is the zone between red and blue lines in the interval the zone [-1.035, 0.948]. For this laminar region we find the closer to power-law distribution of laminar lengths (b) The laminar lengths distribution produced inside the grass laminar region. The red line is the fitting function of eq.7. The results of fitting are: $p_2$=1.34, $p_3$=0, $R^2$ =0.995. These results agree with the CM model as well as the relaxation interval in biological membrane .*

## Section7. Conclusions

In this work, we have shown that a form of HANN whose operation is based on concepts of physics, such as critical intermittency - tricritical intermittency, can produce a temporal spike train that belongs to the category of biological spikes. Given that the biological spike carries information in a biological neural network and therefore behaves as a "quantum" of thought, avenues are opened for future research where ANN will be equipped with properties that are close to physical reality, i.e. the biological brain.

# Bibliografy.

**1.** Hodgkin AL, Huxley AF, Katz B (April 1952). *"Measurement of current-voltage relations in the membrane of the giant axon of Loligo". The Journal of Physiology. **116** (4): 424–48. doi:10.1113/jphysiol.1952.sp004716. PMC 1392219. PMID 14946712.*

**2.** Kosmidis EK, Contoyiannis YF, Papatheodoropoulos C, Diakonos FK. Traits of criticality inmembrane potential fluctuations of pyramidal neurons in the CA1 region of rat hippocampus. Eur J Neurosci. 2018;48:235–2343. https://doi.org/10.1111/ejn. 14117.

**3.** Yiannis F. Contoyiannis ,Efstratios K. Kosmidis, Fotios K. Diakonos, Myron Kampitakis, Stelios M. Potirakis**.** A hybrid artificial neural network for the generation of critical fluctuations and inter -spike intervals..Chaos, Solitons and Fractals. 159. 2022.

4.. Maass W (1997). *"Networks of spiking neurons: The third generation of neural network models". Neural Networks. **10** (9): 1659–1671. doi:10.1016/S0893-6080(97)00011-7. ISSN 0893-6080*

5. Yamazaki, Kashu; Vo-Ho, Viet-Khoa; Bulsara, Darshan; Le, Ngan (July 2022). *"Spiking Neural Networks and Their Applications: A*

*Review"*. *Brain Sciences.* **12** (7): 863.
doi:10.3390/brainsci12070863. ISSN 2076-3425. PMC 9313413.
PMID 35884670

6.      Ribeiro, Bernardete; Antunes, Francisco; Perdigão, Dylan; Silva,
Catarina (2024-08-05). "Convolutional Spiking Neural Networks
targeting learning and inference in highly imbalanced datasets". Pattern
Recognition Letters. doi:10.104..

7.      Lee D, Lee G, Kwon D, Lee S, Kim Y, Kim J (June 2018).
"Flexon: A Flexible Digital Neuron for Efficient Spiking Neural Network
Simulations". 2018 ACM/IEEE 45th Annual International Symposium on
Computer Architecture (ISCA). pp. 275–288.
doi:10.1109/isca.2018.00032. ISBN 978-1-5386-5984-7. S2CID
50778421.

8. Ganguly, Chittotosh; Bezugam, Sai Sukruth; Abs, Elisabeth; Payvand,
Melika; Dey, Sounak; Suri, Manan (2024-02-01). "Spike frequency
adaptation: bridging neural models and neuromorphic applications".
Communications Engineering. 3 (1): 22. doi:10.1038/s44172-024-00165-
9. ISSN 2731-3395. PMC 1105316016/j.patrec.2024.08.002.

9. Gerstner W (2001). "Spiking Neurons". In Maass W, Bishop CM
(eds.). Pulsed Neural Networks. MIT Press. ISBN 978-0-262-63221-8.

10. Gerstner W, Kistler WM (2002). Spiking neuron models: single
neurons, populations, plasticity. Cambridge, U.K.: Cambridge University
Press. ISBN 0-511-07817-X. OCLC 57417395.

11. De Wilde P. Neural network models. 2nd ed. Springer; 1997.

12. SpikeProp: backpropagation for networks of spiking neurons.
Sander M. Bohte , Joost N. Kok Johannes A. La Poutré Conference:
ESANN 2000, 8th European Symposium on Artificial Neural
Networks, Bruges, Belgium, April 26-28, 2000, Proceedings.

13. *Furber, Steve (August 2016). "Large-scale neuromorphic computing systems". Journal of Neural Engineering. **13** (5): 051001. Bibcode:2016JNEng..13e1001F. doi:10.1088/1741-2560/13/5/051001. ISSN 1741-2552. PMID 27529195.*

14. Ballard, D. H. (1987, July). Modular learning in neural networks. In Proceedings of the sixth National conference on Artificial intelligence-Volume 1 (pp. 279-284).

15.    Michael Pfeiffer[*]    Thomas Pfeil Deep Learning With Spiking Neurons: Opportunities and Challenges. https://doi.org/10.3389/fnins.2018.00774

16. *Tavanaei A, Ghodrati M, Kheradpisheh SR, Masquelier T, Maida A (March 2019). "Deep learning in spiking neural networks". Neural Networks. **111**: 47–63. arXiv:1804.08150. doi:10.1016/j.neunet.2018.12.002. PMID 30682710. S2CID 5039751.*

17. Furber, Steve (August 2016). "Large-scale neuromorphic computing systems". Journal of Neural Engineering. 13 (5): 051001. Bibcode:2016JNEng..13e1001F. doi:10.1088/1741-2560/13/5/051001. ISSN 1741-2552. PMID 27529195.

18. Stelios M Potirakis, Fotios K Diakonos, Yiannis F Contoyiannis. A Spike Train Production Mechanism Based on Intermittency Dynamics. Entropy, 27(3)  , 2025.

19. Intermittent Dynamics of Critical Fluctuations
Contoyiannis, Y.F., Diakonos, F.K., Malakis, A.
Physical Review Letters, 2002, 89(3).

20. Huang K 1987 Statistical Mechanics 2nd (New York: Wiley)

21. Tricritical crossover in earthquake preparation by analyzing preseismic electromagnetic emissions
Contoyiannis, Y., Potirakis, S.M., Eftaxias, K., Contoyianni, L.
Journal of Geodynamics, 2015, 84, pp. 40–54.

22. H. G. Schuster, *Deterministic Chaos* (VCH, Weinheim, 1998).

23. Yiannis Contoyiannis and Stelios M Potirakis J. Signatures of the symmetry breaking phenomenon in pre-seismic electromagnetic emissions. Stat. Mech. (2018) 083208.

24. Contoyiannis, Y.F.; Potirakis, S.M.; Diakonos, F.K.; Kosmidis, E.K. Criticality in a hybrid spin model with Fermi-Dirac statistics. *Phys. A* **2021**, *577*, 126073. https://doi.org/10.1016/j.physa.2021.126073.

25. N. Metropolis, A.W. Rosenbluth, M.N. Rosenbluth, A.H. Teller, E. Teller, J. Chem. Phys. 21 (1953) 1087, http://dx.doi.org/10.1063/1.1699114.

26. R. Rubin, L.F. Abbott, H. Sompolinsky, Proc. Natl. Acad. Sci. USA 114 (2017) 9366–9375, http://dx.doi.org/10.1073/pnas.1705841114.

27. L.D. Landau, E.M. Lifshitz, Statistical Physics Part 2, Course of Theoretical Physics Volume 9, Pergamon Press, Oxford, 1969.

28. L. Witthauer, M. Dieterle, The Phase Transition of the 2D-Ising Model, 2007, http://quantumtheory.physik.unibas.ch/bruder/Semesterprojekte2007/p1/index.htmlx1-110002.1.6